\documentclass[twocolumn,prl,showpacs]{revtex4-1}
\usepackage{graphicx}
\usepackage{color}
\begin{document}
\title{Base Pair Openings and Temperature Dependence of DNA Flexibility}
\author{Nikos Theodorakopoulos$^{1,2,3}$ and 
Michel Peyrard$^{3}$
} 
\affiliation{
$^{1}$Theoretical and Physical Chemistry Institute, National Hellenic Research
Foundation,\\ 
Vasileos Constantinou 48, 116 35 Athens, Greece \\
$^{2}$Fachbereich Physik, Universit\"at Konstanz, 78457 Konstanz, Germany\\
$^{3}$Ecole 
Normale Sup\'erieure de Lyon, Laboratoire de Physique CNRS,
46 all\'ee d'Italie, 69364 Lyon Cedex 7, France
}
\date{\today}
%\draft
\begin{abstract}
The relationship of base pair openings to DNA flexibility is examined.  
Published experimental data on the temperature dependence of the
persistence length by two different groups 
are well described  in terms of an inhomogeneous Kratky-Porot model with
soft and hard joints, 
corresponding to open and closed base pairs, and sequence-dependent
statistical information 
about the state of each pair provided by a   
Peyrard-Bishop-Dauxois (PBD) model calculation 
with no freely adjustable parameters. 
\end{abstract}
\maketitle
%87.10.-e General theory and mathematical aspects
%87.14.-g Biomolecules: types
%87.14.gk DNA 
%87.15.Zg Phase transitions
%82.35.Lr Physical properties of polymers (82: Physical chemistry,
%         chemical phys)
%87.10.Pq Elasticity theory (ni Biological Physics)
%87.15.-v Biomolecules: structure and physical properties
%%%%%%%%%%%%%%%%%%%%%%%%%%%%%%%%

The bending flexibility of DNA is essential in key biological processes, such
as its packing in chromatin or in viruses. While single strands of
DNA (or RNA) are highly flexible owing  to an almost free rotation
around single chemical bonds, the double helical structure  with
its hydrogen-bonded stacked base pairs favors a very rigid
configuration. One may ask what are the natural limits of this rigidity.
% Do they arise from  the finite strength of the network of
% chemical bonds that make the double helix or are other mechanisms
% involving e.g. sequence and environmental factors contributing? 
This question has been addressed almost since the discovery of DNA
structure and has led to controversial results.  

A measure of the flexibility of the molecule is provided by its {\em persistence
  length}, which is the correlation length for the thermal fluctuations
of a vector tangent to the helix axis. Early measurements were made by light
scattering methods \cite{Peterlin}. The temperature dependence of
persistence length 
was measured by Gray and Hearst \cite{Gray} in sedimentation experiments.
The analysis of these results 
by a model that describes DNA as a flexible rod, the 
so-called wormlike chain (WLC) model, led the authors to
suggest that a statistically homogeneous WLC description did not apply,
and to suggest that, instead of a smooth bending, the
flexibility could come from sharp bends, at ``flexible
joints''. This appeared consistent with the observation of ``DNA
breathing'' by von Hippel  and coworkers \cite{vonHippel}. A 
denaturation of the double helix, forming local single strands, would
provide those ``joints''. Another hint of the possibility
of sharp bends came from cyclization measurements of short 
chains \cite{Cloutier} which gave a much higher probability to find the
two ends of a short DNA molecule next to each other than predicted by
a homogeneous WLC model leading to smooth bendings. A theoretical
analysis \cite{YanMarko} showed that a localized single-stranded bubble
mechanism was a possible explanation, compatible with the energetics of
the thermal fluctuations of DNA. However, the interpretation of
cyclization experiments was 
criticized in later work \cite{DuVologodskii} identifying 
the high ligase protein used in \cite{Cloutier} as the source of a
possible artefact. 

As the local opening of DNA can be thermally induced, a study of the
temperature dependence of the persistence length should provide a
clue. In fact recent 
measurements \cite{Volo} performed between 5 and 60 C reveal  
a much stronger variation of the persistence length than that either found
in earlier work \cite{Gray} or predicted by a WLC model with a fixed
stiffness constant. 
The authors of Ref.~\onlinecite{Volo} ruled out the possibility that local
openings could be responsible for this behavior on the basis of the low opening
probabilities derived from 
%imino proton exchange studies
% ; the latter experiments detect the rate of
% proton-deuterium exchange  
the rate of exchange of the protons involved in the base-pair hydrogen
bonds when DNA is in 
solution in deuterated water \cite{Gueron87}.

A promising experimental route to resolve the issue of whether local
openings are relevant to increased DNA flexibility  
would be to examine the strongly fluctuating premelting and 
melting regimes.  It turns out that this was done many years ago in a pioneering
study of magnetic birefringence of DNA solution by Maret and coworkers
\cite{Maret75}.  In this work we interpret the results of \cite{Maret75}
in terms of base-pair openings whose populations can be determined by a
lattice-dynamically motivated mesoscopic model of  DNA.  Furthermore, we
apply these ideas to a direct calculation of the average  end-to-end
distance 
which allows us to compare predicted effective persistence length with the
results of \cite{Volo}. Our analysis of the two experiments in terms of
standard model ideas of heterogenous polymer elasticity and DNA bubbles
describes the full temperature dependence of the two experimental
results and suggests that a quantitative link
between base pair openings and local flexibility can be detected in measured
properties of DNA solutions. 

We model the elastic behavior of a DNA chain of  $N$ base pairs in a magnetic
field $B$ along the $z$ axis in terms of a heterogeneous Kratky-Porod
chain \cite{KP}

\begin{equation}
	H_{KP} = -\sum_{j=1}^{N-1} J_{j,j+1}{\vec S}_{j} \cdot {\vec S}_{j+1}
        -  \frac{1}{2}\Delta \chi  
\sum_{j=1}^{N} (B  S_j^z ) ^2  
\label{eq:KPHam}
\end{equation}
\begin{figure}[h]
\vskip -.75truecm
\includegraphics[width=0.4\textwidth,height=0.4\textwidth]{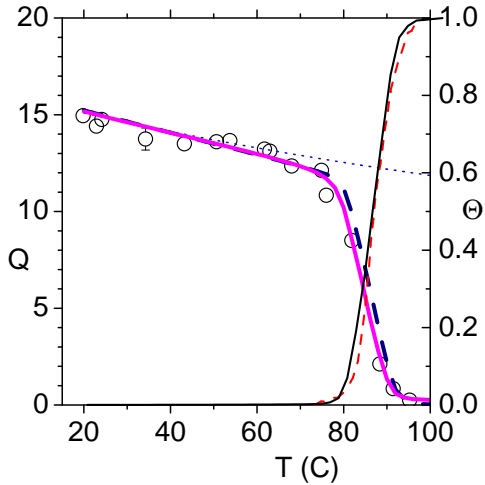}	
\vskip -0.5truecm
\caption{   (color online)
{\small  The quantity Q (left y-axis), as extracted from measurements of
  magnetic birefringence (circles, adapted from \cite{Maret75}), and
  calculated (i) on the basis of Eq. \ref{eq:Qcluster}  (dashed curve), (ii)
  from the heterogeneous KP model (full line); the dotted curve shows a
  $1/T$ behavior for 
  comparison. Also shown (right y-axis) is the measured melting fraction
  for a similar 
  sample \cite{MarmurDoty} (dotted line) and as calculated from a $290 kb$
  sequence of {\em bos taurus} \cite{BovSeq} (thin full line). 
}} 
\label{fig:biref}
\end{figure}
where ${\vec S}_j$ is the unit vector describing the direction of the $j$th
base-pair plane.
The first term is simply a discrete version of the WLC model, with a
local stiffness constant $ J_{j,j+1}$. The second term describes the
contribution of the magnetic energy.
$\Delta \chi = \chi_{||}-\chi_{\perp}$ is the anisotropy
of the magnetic susceptibility of a single base pair, known to be negative
(diamagnetic) in a fixed molecular frame where the parallel direction is along
the axis of the helix. The $(S_j^z)^ 2 \equiv \cos ^2 \theta_j$ factor follows
from the Euler $(\theta, \phi)$ rotation which transforms
the susceptibility tensor to the laboratory frame if one integrates over all
possible azimuthal angles (axial symmetry). 
The stiffness constants will be taken as either $J$ (stiff) or $J'$ (soft)
with an appropriate probability distribution (cf  below).  

The magnetically induced optical birefringence 
$\Delta n = n_{||}-n_{\perp}$ of the
refractive index along directions parallel and perpendicular to the
field, measured in \cite{Maret75}, is
produced by the difference in the (optical frequency) dielectric constant
which originates in  the (electrically) anisotropically polarizable units, the
DNA base pairs. Standard electromagnetic theory implies 
\begin{equation}
	\Delta n = \frac{2\pi} {n} \> \rho \> \Delta \alpha \> {\hat Q}
\end{equation}
where $n$ is the average index of refraction, $\Delta \alpha = \alpha_{||}-
\alpha_{\perp}$ the anisotropy of the electronic polarizability tensor of a
single base pair in the molecular frame, $\rho$ the number of monomers (base
pairs) per unit volume, and  
\begin{equation}
	{\hat Q} = \frac{1}{N}\sum_{j=1}^{N}\frac{1}{2} 
        \langle(3\cos^ 2\theta_j-1) \rangle
\end{equation}
is an averaged orientational factor whose origin can be traced, in analogy
with the second term in the magnetic Hamiltonian (\ref{eq:KPHam}), to the
anisotropy of  the axially symmetric polarizability tensor in the laboratory
frame.  Following \cite{MaretWeill} we consider any local field corrections to
be incorporated in effective values of the polarizability. The thermal
averaging is performed over conformations of the molecule according to
(\ref{eq:KPHam}). Since, even for high magnetic fields, $\Delta \chi B^2$ is
much smaller than the typical thermal energies $k_B T$, the averaging of the
orientational factor can be evaluated to leading order in $B^2$ 
\begin{eqnarray}
\nonumber
	{\hat Q} &= &\Delta \chi \>\frac{B^2}{2k_B T}\> Q\quad,\quad {\rm
          where}\\ 
\nonumber
Q &=&  \frac{1}{N}\sum_{i,j=1}^{N} \langle \frac{1}{2}(3\cos^ 2\theta_j-1) \cos^2
\theta_i \rangle_0 
\label{eq:AvS}
\end{eqnarray}
and the $0$ subscript implies averaging over field-free configurations of the
KP chain (\ref{eq:KPHam}). For a homogeneous KP chain of $\nu$ monomers
the value of $Q$ is \cite{Wilson} 
\begin{equation}
	 Q_\nu= \frac{2}{15} \left\{
\nu \frac{1+u}{1-u} - 2\frac{u(1-u^\nu)}{(1-u)^2}
\right\}
\label{eq:Qhom}
\end{equation}
where $u=1-(3/K)(\coth K-1/K)$, $K=J/k_B T$.
For the case of DNA, the orientational averages are computed 
 in an heterogeneous KP ensemble where the stiffness constants 
depend on whether the molecule is locally in a double-stranded or
open configuration
\begin{equation}
	 J_{j,j+1}= (1-P_0^{j,j+1}) J + P_0^{j,j+1} J' \; .
\label{eq:AvJ}
\end{equation}
$P_0^{j,j+1}$ is  the joint probability that base pairs $(j)$ and
$(j+1)$ are open. It is obtained from an exact calculation
of the partition function of the
mesoscopic PBD  Hamiltonian \cite{PBD}, which associates a transverse
coordinate $y_n$ with each base pair and is known to describe the thermal   
denaturation of long DNA chains with known sequence and salt content,
with the method described in Ref.\ \cite{NTh2011}. The model
parameters, determined in earlier analysis of melting profiles
\cite{NTh2010} have {\em not} been adjusted for this study.

Fig.~\ref{fig:biref} compares theoretical predictions as outlined above with
the experimental data on calf thymus DNA. The experimental
values of the dimensionless quantity $Q$ have
been extracted from the measured birefringence values \cite{Maret75} 
using a mean index of refraction $n=1.33$ \cite{MaretWeill}, 
and the values $\Delta \chi=-1.55 \times 10^{-20}$ erg/$T^2$
\cite{chiDNA} and $\Delta 
\alpha=-18.2\;$\AA$^3$ \cite{Stellwagen} for the anisotropies,
respectively, of the 
diamagnetic susceptibility and electronic polarizability per base
pair. The parameter $J$ has been set to
$J=7.02 \times 10^{-12}$ erg
which corresponds  to a room temperature persistence length of  $59 $ nm.
Consistently with the parameters of the PBD model \cite{NTh2010}  
which introduce a
ratio of $50$ between the base pair interactions for stacked or
unstacked bases, we set $J' =J/50  =  0.14 \times 10^{-12}\;$erg,
which gives a persistence length of $1.87$ nm for single-stranded DNA.
For reference purposes, we show on Fig.~\ref{fig:biref} the
experimental calf thymus
melting profile \cite{MarmurDoty} and a PBD-model computed profile from a
$290$ kbp long segment of bovine chromosome with a similar melting temperature
\cite{BovSeq}.  The computed profile has been obtained with the model
parameters and 
the numerical procedure described in \cite{NTh2010}. 
 
\begin{figure}[t]
\vskip -.5truecm
\includegraphics[width=0.4\textwidth,height=0.4\textwidth]%
{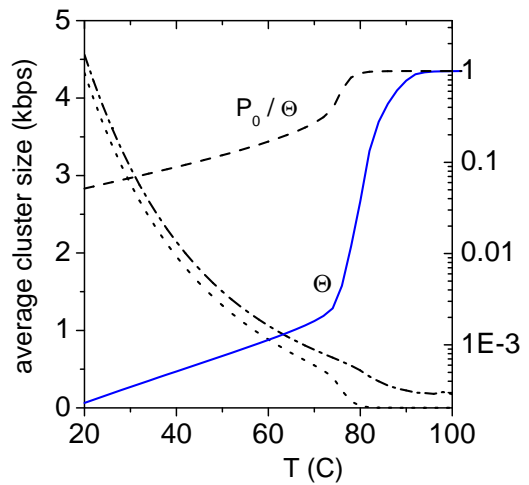}
\vskip -0.5truecm
\caption{ (color online)
{\small % {\it left panel:} 
Some average properties of the bovine sequence \cite{BovSeq}. 
Left $y$-scale: the average size (in bps) of intact clusters (dash-dotted
curve); the inverse of the melting fraction (dotted curve). Right $y$-scale:
the melting fraction $\Theta$ (solid curve) and the conditional probability
$P_0/\Theta$ that if a base pair is in the open state, the next one will also
be in the open state (dashed curve); note that the latter quantity is much
larger than $\Theta$ (bubble aggregation tendency).  
}}
\label{fig:bubbcorr}
%\vskip -0.5truecm
\end{figure}

To complete this study 
we would like to present an alternative way of computing the dimensionless
quantity $Q$ using PBD model input on base pair openings, following the
analysis used recently in the interpretation of neutron scattering experiments
on B-DNA \cite{Wildes}. Although slightly less accurate than the
previous numeric evaluation, the alternative route illustrates some of the
underlying properties of open vs. closed base pairs.  
According to that picture, neutron diffraction - and in the present case
magnetic birefringence - is produced from oriented intact clusters of bound
base pairs. An intact cluster of length $\nu$ (defined as enclosed by open
base pairs on both sides) occurs with a probability $P_\nu$ which satisfies
the sum rules \cite{NTh2011}  
% \begin{equation}
$ \sum_{\nu=0}^{\infty} P_\nu =   \Theta $, % \qquad
$ \sum_{\nu=1}^{\infty}\nu P_\nu =   1- \Theta$,  % \quad.
% \label{eq:sumrules}
% \end{equation}
where $\Theta$ is the overall fraction of open base pairs. It follows that the
average size of intact clusters (measured in base pairs) 
\begin{equation}
s = \frac{\sum_{\nu=1}^{\infty}\nu P_\nu}{\sum_{\nu=1}^{\infty}
  P_\nu}=\frac{1-\Theta}{\Theta-P_0} 
\end{equation}
can be determined in terms of $\Theta$ and the \lq\lq zero size cluster
probability" $P_0$ (cf. Fig.~\ref{fig:bubbcorr}); note that the latter
quantity is in fact the joint probability of two successive base pairs being
in the open state. Moreover, the $P_\nu$'s behave exponentially, i.e., 
$ P_\nu = A e^{-\sigma \nu}, \nu \geq 1$, 
where $A=(1-\Theta)/[s(s-1)]$ and $\sigma = -\ln(1-1/s)$,  to a high degree of
accuracy \cite{Wildes, NTh2011}.  

It is therefore possible, in an approximate sense which neglects contributions
to birefringence from open base pairs, to evaluate $Q$ as an average over all
intact clusters, treated as if they were freely jointed
\begin{equation}
%\nonumber
Q = \sum_{\nu=1}^{\infty} P_\nu Q_\nu  
 \sim \frac{4}{45} K (1-\Theta) \frac{1}{1+ ({K}/{3s})}
\label{eq:Qcluster}
\end{equation}
where it should be recognized that for typical ds-DNA parameters, $K>>1$ is
very nearly equal to the persistence length $\lambda$ expressed in units of
the monomer distance $a=0.34$ nm.  The dashed curve in Fig.~\ref{fig:biref}
expresses Eq. \ref{eq:Qcluster} for the same $J=7.02 \times 10^{-12}$
erg as above. 
%
%
%\textcolor{red}{
The agreement between the theoretical predictions and the magnetic
birefringence experiments (Fig.~\ref{fig:biref}), in the whole
temperature range from room to melting temperatures, validates this
approach which implies that fluctuational openings and persistence
length of DNA are intimately linked, even at temperatures significantly
below melting.%}

\begin{figure}[t]
\vskip -.5truecm
%\resizebox{0.3\textwidth}{!}
%\resizebox{0.7\textwidth}{!}
\includegraphics[width=0.4\textwidth,height=0.4\textwidth]{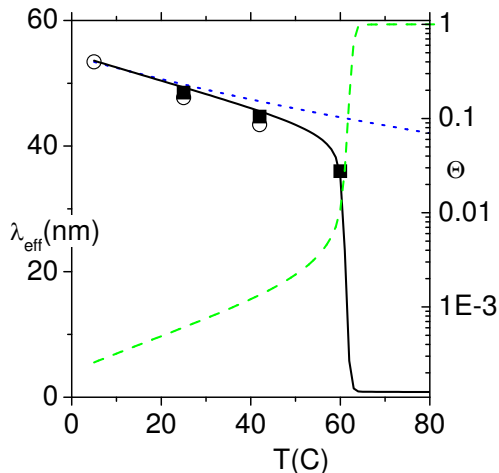}
\vskip -0.5truecm
\caption{ (color online)
{\small % {\it left panel:} 
Persistence length obtained by cyclization of DNA fragments (filled squares,
open circles, redrawn from ref. \cite{Volo} ). The solid curve shows the
estimate obtained from the inhomogeneous KP chain with stiff and soft joints
distributed according to (\ref{eq:AvJ}) and probabilities from a PBD
calculation.   
Also shown is a  $1/T$ reference curve (dotted line), and the melting
fraction (dashed line, right 
y-axis) obtained from the PBD calculation. 
}}
\label{fig:CyclData}
\vskip -0.2truecm
\end{figure}

We now turn our attention to the more recent set of persistence length
measurements which were based on cyclization of $\lambda $-phage DNA
fragments \cite{Volo}. 
The overall temperature dependence of the persistence length
appears (Fig.~\ref{fig:CyclData}) to be stronger than the $1/T$ predicted by
the homogeneous KP model. In what follows we will show that such
a strong temperature dependence is entirely consistent with the
bubble-based scenario proposed above for the description of the
birefringence data. We consider a heterogeneous KP model with 
stiffness constants $J$ or $J'$ determined
with the probabilities given by the PBD model (using Eq.~ (\ref{eq:AvJ}),
$J'=0.14 \times 10^{-12}$ erg and slightly 
varying $J$). It can be used to compute a theoretical end-to-end
distance, based on the bubble-based scenario, 
for the 200 bp sequence of the $\lambda -$phage 
for a Na$^+$
concentration of $0.004\;$M (cf. \cite{Volo}), from
\begin{equation}
  \langle R^2 \rangle = \frac{1}{N}\sum_{i,j=1}^{N} \langle {\vec S}_i \cdot  {\vec
  S}_j \rangle \; .
\end{equation}
A subsequent analysis based on the homogeneous WLC model as in
\cite{Volo,Volo2002}, with the same total length $L=Na$, i.e.\ with 
\begin{equation}
   \langle R^2 \rangle =  2 L \lambda - 2  \lambda^2(1-e^{-L/\lambda})
\end{equation}
allows us to extract an effective persistence length which can be
directly compared with the experimental data. The resulting
effective persistence length, obtained for $J=6.13 \times 10^{-12} $
erg,
plotted in Fig.~\ref{fig:CyclData}, shows that %\textcolor{red}{
the data   of \cite{Volo} are consistent with a bubble-based analysis.%}
\par
%
%\bigskip
We have shown that the results of the two experiments which measured DNA
persistence length in a temperature range sufficiently 
broad to detect deviations from a $1/T$ behavior can be fitted by adopting a
{\em single} 
approach combining %the Kratky-Porod model,i.e.\ 
a discrete version of the WLC model %generally used, 
with the statistical physics of the base-pair fluctuational openings. 
Three comments are in order here. 

%\textcolor{red}{
First, models which do not include fluctuational
  openings lead to a persistence length that varies as $1/T$, in significant
  disagreement with experiments. The data of Ref.~\cite{Volo} show that,
  in the $5$ -- $42^{\circ}$C range the observed variation is $1.45$
  times bigger than predicted by an $1/T$ law.%}

The second comment concerns compatibility of our results with 
the low densities ($<10^{-5}$) of fluctuational openings and the
associated long lifetimes ($\sim 10$ msec) of base pairs  
measured by imino proton exchange  \cite{Gueron87}. The results of
Ref. \cite{Gueron87} 
show very clearly that the activation enthalpy for the
opening, %, which is around $45\;$kJ/mol in the $0-25^{\circ}$C range,
which stays roughly constant in the $0-25^{\circ}$C range,
increases significantly above $25^{\circ}$C. In other words there is an
acceleration of the exponential rate at which lifetimes decrease with
increasing temperature. A straightforward extrapolation of data
appearing  in Fig.~12 of [\onlinecite{Gueron87}b] suggests an increase
%\textcolor{red}{
by a factor of 10 
between $25^{\circ}$C and $40^{\circ}$C. Thus, events considered as
exceptional at room temperature become more plausible at
a temperature close to physiological temperature.%}
Moreover it should be noted that relatively few
disruptions are sufficient to significantly change the persistence
length. 
%% Other experiments have pointed out that the influence of large
%% fluctuations in AT-rich regions starts to show up around  $40^{\circ}$C
%% and extends to some distance \cite{CuestaNAR}. 
Perhaps the most striking indirect evidence of the role of small numbers
of fluctuational openings in the premelting regime of genomic DNA comes
from viscosity measurements \cite{Freund63}.  

The third comment refers to possible extensions and applications of the
present work. %\textcolor{red}{
The approach used here, owing to its computational efficiency, can be exploited in its present form for a systematic study of sequence effects on local flexibility \cite{WeberFlex}. On the other hand, it is likely that extended soft regions of the DNA molecule may require a less minimalist approach than the one presented here. In particular, it is possible that a more complex interplay between the  torsional, bending, and local melting \cite{Chen} effects may determine the details of the premelting regime of cyclization experiments; as more experimental data become available, generalizations of the present work in that direction may become fruitful.%} 

As a final remark, it may be noted that the quantitative link between
fluctuational openings and local flexibility explored in this work
provides further evidence that the dynamical character of DNA should not
be underestimated, even away from its melting transition. 

%\bigskip
N. T. acknowledges support by the program Accueil-Pro of
R\'egion Rh\^one-Alpes. 

%%%%%%%%%%%%%%%%%%%

\end{document}